\documentclass[fleqn,twoside]{article}
\usepackage{espcrc2}
\def\g2{{$(g-2)$} }

\usepackage{graphicx}
\usepackage[figuresright]{rotating}

\newcommand{\AmS}{{\protect\the\textfont2
  A\kern-.1667em\lower.5ex\hbox{M}\kern-.125emS}}

\title{Future Muon Dipole Moment Measurements}

\author{B. Lee Roberts\thanks{This work supported in part by the 
        U.S. National Science Foundation}\\
Department of Physics, Boston University
         \\ 
        590 Commonwealth Avenue \\ Boston, MA 02215 USA%
        }
       
\begin{document}

\begin{abstract}
From the famous experiments of Stern and Gerlach to the present,
measurements of magnetic dipole moments, and searches for electric
dipole moments of ``elementary'' particles have played a major
role in our understanding of sub-atomic physics.  In this talk
I discuss the progress on measurements and theory of the 
magnetic dipole moment of the  muon.  I also discuss
a new proposal to search for a permanent electric dipole moment (EDM) of
the muon and put it into the more general context of other EDM searches.
These experiments, along with searches for the lepton flavor violating
decays $\mu \rightarrow e \gamma$ and $\mu^- + A \rightarrow e^- + A$, 
provide a path to the high-energy frontier through precision measurements.
\vspace{1pc}
\end{abstract}

\maketitle

\section{Introduction and theory of the lepton anomalies}

Over the past 83 years, the study of dipole moments of 
elementary particles has provided a wealth of information on 
subatomic physics.  From the pioneering work of Stern\cite{stern}
through the discovery of the large anomalous magnetic moments
of the proton\cite{sternp} and neutron\cite{nmdm}, the ground
work was laid for the discovery of spin, of radiative corrections
and the renormalizable theory of QED, of the quark structure of
baryons and the development of QCD.  

A charged particle with spin $\vec s$ has a magnetic moment
\begin{equation}
 \vec \mu_s = g_s ( {e \over 2m} ) \vec s;
\   a \equiv { (g_s -2) \over 2};\   \mu = (1 + a){e \hbar \over 2m};
\end{equation}
where $g_s$ is the gyromagnetic ratio, $a$ is the anomaly,
and the latter expression is what one finds in the Particle
Data Tables.\cite{pdg}

\begin{figure}[h!]
  \includegraphics[width=0.45\textwidth]{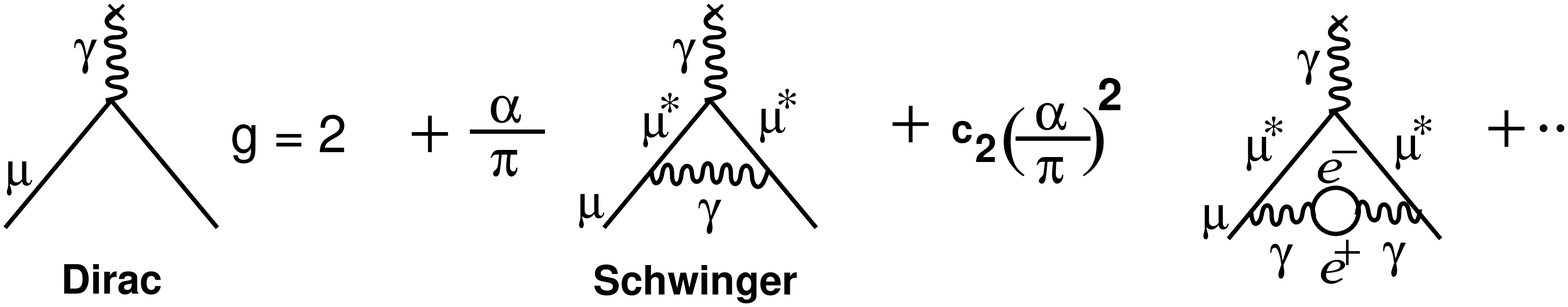}
  \caption{A schematic of the first few terms in the QED expansion
for the muon. The vacuum polarization term shown is one of
five of order $(\alpha/\pi)^2$.}
\label{fg:aexpan}
\end{figure}

The Dirac equation tells us that for spin one-half point-like particles,
 $g\equiv 2 $ for spin angular momentum,
and is unity for orbital angular momentum (the
latter having been verified experimentally\cite{kusch}).  
For point particles, the anomaly arises from radiative corrections,
two examples of which are shown in Fig.~\ref{fg:aexpan}.
The  lowest-order correction gives the famous 
Schwinger\cite{schwinger}
result,  $a =\alpha/2 \pi$,
which was verified experimentally
by Foley and Kusch.\cite{kusch}
The situation for baryons is quite different, since their
internal quark structure gives them large anomalies.

In general {$a$} (or $g$) is an expansion 
in $ \left({\alpha \over  \pi}\right)$,
\begin{equation}
a = C_1\ \left( {\alpha \over  \pi}\right) 
+ C_2\  \left( {\alpha \over  \pi}\right)^2
+C_3\ \left( {\alpha \over  \pi}\right)^3 
+ C_4\ \left( {\alpha \over  \pi}\right)^4 + \cdots
\end{equation}
with 1 diagram for the Schwinger (second-order) contribution,
5 for the fourth order, 40 for the sixth order, 891 for the eighth
order.

The QED contributions to electron and muon \g2 have now been calculated
through eighth order, $(\alpha/\pi)^4$ and the 
tenth-order contribution has been estimated.\cite{kinqed}

\begin{table}[h!]
\begin{center}
\begin{tabular}{cccc} \hline
      &{$\vec E$ }
      &{$\vec B$ }
      & {$\vec \mu$ or  $\vec d$} \\
\hline
{\sl P} & - & + & + \\
{\sl C} & - & - & - \\
{\sl T} & + & - & - \\
\hline
\end{tabular}
\caption{Transformation properties of the magnetic and electric fields and
dipole moments.}
\label{tb:tranprop}
\end{center}
\end{table}

While magnetic dipole 
moments (MDMs) are a natural property of charged particles with spin,
electric dipole moments (EDMs) are forbidden both by parity {\sl (P)} and 
by time reversal {\sl (T)} symmetries.\cite{pr,landau,ramsey}
 This can be seen by examining
the Hamiltonian for a spin one-half particle in the presence of
both an electric and magnetic field,
${\mathcal H} = - \vec \mu \cdot \vec B  - \vec d \cdot \vec E$.
The transformation properties of $\vec E$, $\vec B$, $\vec \mu$ and $\vec d$
are given in Table \ref{tb:tranprop}, and we see that while
$\vec \mu \cdot \vec B$ is even under all three,
$\vec d \cdot \vec E$ is odd under both {\sl P} and
{\sl T}.  Thus
the existence of an EDM implies that both {\sl P} and {\sl T} are violated.
In the context of {\sl CPT} symmetry, an EDM implies {\sl CP} violation.
The standard model value for the
electron and muon EDMs are well beyond the reach of 
experiment (see Table \ref{tb:edm}), so
observation of a non-zero $e$ or $\mu$ EDM
would be a clear signal for new physics.
 Since the presently known {\sl CP} violation
is inadequate to describe the baryon asymmetry in the universe, additional
sources of {\sl CP} violation should be present. Furthermore, we do
expect to find {\sl CP} violation in the lepton sector.  New dynamics
such as supersymmetry could easily produce new sources of {\sl CP} violation
which could have a 
possible connection with cosmology (leptogenesis).\cite{ellis1,ellis2,bdm}

\begin{table}[h!]
\begin{tabular}{ccc} 
\hline
{ Particle}  & { Present Limit} & { SM Value} \\
             &  { (e-cm) } & { (e-cm) }\\
\hline

 n\cite{nedm} &{$6.3 \times 10^{-26}$ } & {$<10^{-31}$ }  \\
\hline
 $e^-$\cite{eedm}  & {$\sim 1.6 \times 10^{-27 }$} & {$<10^{-38}$ } \\
\hline
 {$\mu$}\cite{cern3} &{$<10^{-18}$ } (CERN) & {$<10^{-35}$ }\\
 & $\sim10^{-19}$ (E821)$^*$\ \ \   & \\
 & {$\sim10^{-24}$ }\ J-PARC$^{\dag}$ \\
\hline
\end{tabular}
\caption{Measured limits on electric dipole moments, and their standard
model values.\hfill\break 
$^*$ Estimated limit, work in progress.\hfill\break
$^{\dag}$Letter of Intent (LOI) to J-PARC for a 
new \break dedicated experiment.\cite{loi}
}
\label{tb:edm}
\end{table}

The  magnetic and electric dipole moments can be represented
as the real and imaginary parts of a generalized dipole operator $D$,
and the interaction Lagrangian becomes
\begin{equation}
{\mathcal{L}}_{dm} = {1\over 2} \left[ D\bar \mu \sigma^{\alpha \beta}
{1 + \gamma_5 \over 2} \  + \ D^* \bar \mu \sigma^{\alpha \beta}
{1 - \gamma_5 \over 2}\right] \mu F_{\alpha \beta}
\end{equation}
with
${\rm Re}\ D\  =\  a_{\mu} {e \over 2 m_{\mu}}$  and
${\rm Im}\ D\  =\  d_{\mu}$.

The electron anomaly is now measured to a relative precision
of about four parts in a 
billion (ppb),\cite{eg2} which is better than the precision on 
the fine-structure constant $\alpha$, and
Kinoshita has used the measured 
electron anomaly to give the best determination
of $\alpha$.\cite{kinalpha} The electron anomaly will be further improved
over the next few years.\cite{gab}
 
The muon anomaly is measured to 0.5 parts per 
million (ppm).\cite{brown2,bennett1,bennett2}  The relative contributions of 
heavier particles to $a$ scales as $(m_e/m_{\mu})^2$, so the muon
has an increased sensitivity to higher mass scale radiative corrections
of about 40,000 over the electron.
At a precision of $\sim 0.5$ ppm, the muon anomaly
is sensitive to $\geq 100$ GeV scale physics.

The standard model value of $a_{\mu}$ has 
measurable contributions from three types of radiative
processes: QED loops containing leptons ($e,\mu,\tau$) and 
photons;\cite{kinqed}
hadronic loops containing hadrons in vacuum polarization 
loops;\cite{davmar,adh,dh,dehz1,dehz2,HMNT,HICHEP,bpp,hk,kn,bcm,mv}
and weak loops involving the $W$ and $Z$ weak gauge bosons
(the standard model Higgs contribution is negligible),\cite{davmar}
\begin{equation}
a_{\mu}{\rm ( SM)}
= a_{\mu}({\rm QED}) + a_{\mu}({\rm Had}) +
a_{\mu}({\rm Weak}).
\end{equation}
A significant 
difference between the experimental value and the standard model
prediction would signify the presence of new physics.
A few examples of such potential contributions are
lepton substructure, anomalous $W-\gamma$
couplings, and supersymmetry.\cite{davmar}

\begin{figure}[h!]
  \includegraphics[width=.45\textwidth]{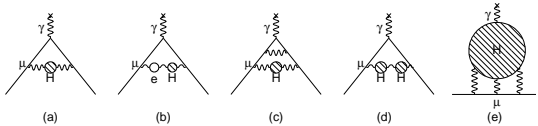}
  \caption{The hadronic contribution to the muon anomaly, where the
dominant contribution comes from (a).  The hadronic light-by-light
contribution is shown in (e).}
\label{fg:had}
\end{figure}

The CERN experiment\cite{cern3}
observed the contribution of hadronic vacuum polarization 
shown in Fig. \ref{fg:had}(a)  at the
8 standard deviation level.  Unfortunately, the hadronic contribution
cannot be calculated directly from QCD, since the energy scale
is very low ($m_{\mu} c^2$), although Blum\cite{blum} has performed 
a proof of principle calculation on the lattice.  
Fortunately dispersion theory 
gives a relationship between the vacuum polarization loop
and the cross section for $e^+ e^- \rightarrow {\rm hadrons}$,
\begin{equation}
a_{\mu}({\rm Had;1})=({\alpha m_{\mu}\over 3\pi})^2
\int^{\infty} _{4m_{\pi}^2} {ds \over s^2}K(s)R(s),
\end{equation}
where
\begin{equation}
R\equiv{ {\sigma_{\rm tot}(e^+e^-\to{\rm hadrons})} \over
\sigma_{\rm tot}(e^+e^-\to\mu^+\mu^-)}
\end{equation}
and experimental
data are used as input.
The factor $s^{-2}$ in the dispersion relation, means that values
of  $R(s)$ at low energies (the $\rho$ resonance) dominate the determination of
$a_{\mu}({\rm Had;1})$. In principle, this information could 
be obtained from hadronic $\tau^-$ decays such as
$\tau^- \rightarrow \pi^- \pi^0 \nu_{\tau} $, which can be 
related to $e^+e^-$ annihilation through the CVC hypothesis and
isospin conservation.\cite{adh,dh,dehz1,dehz2} 
However, inconsistencies between
information obtained from $e^+e^-$ annihilation and hadronic
tau decays, plus an independent 
confirmation of the CMD2  high-precision $e^+e^-$ cross-section
measurements by the KLOE collaboration,\cite{KLOE} 
have prompted Davier, H\"ocker, et al,  to state that 
until these inconsistencies can be understood
only the $e^+e^-$ data should be used to 
determine $a_{\mu}({\rm Had;1})$.\cite{HICHEP}

\begin{figure}[h!]
  \includegraphics[width=.45\textwidth]{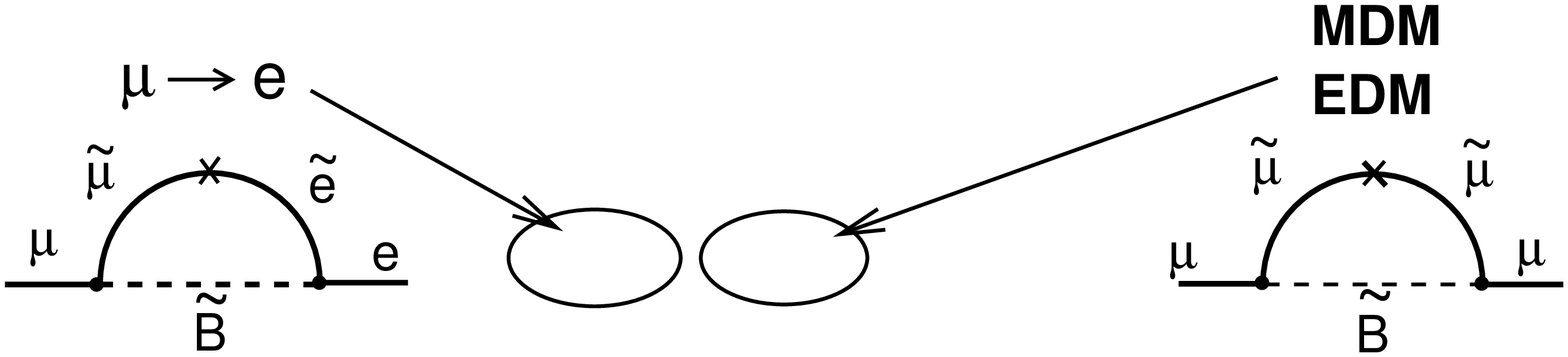}
  \caption{The supersymmetric contributions to the anomaly, and to
$\mu \rightarrow e$ conversion, showing the relevant slepton mixing matrix
elements. The MDM and EDM give the real and imaginary parts of
the matrix element, respectively.}  
\label{fg:susy}
\end{figure}

The hadronic light-by-light contribution (see Fig. \ref{fg:had}(e)) 
has been the topic of much
theoretical investigation.\cite{bpp,hk,kn,bcm,mv}  Unlike the lowest-order
contribution, it can only be calculated from a model, and this contribution
is likely to provide the ultimate limit to the precision of the
standard-model
value of $a_{\mu}$.

One of the very useful roles the measurements of $a_{\mu}$ have played in the 
past is placing serious restrictions on physics beyond the standard model.
With the development of supersymmetric theories as a favored scheme of
physics beyond the standard model, interest in the experimental and
theoretical value of $a_{\mu}$ has grown substantially.  SUSY contributions
to $a_{\mu}$ could be at a measurable level in a broad range of models.
Furthermore, there is a complementarity between the SUSY contributions 
to the MDM, EDM and transition moment for the lepton-flavor
violating (LFV) process $\mu^- \rightarrow e^-$ in the field of a nucleus.  
The MDM and EDM are related to the real and
imaginary parts of the diagonal element of the slepton 
mixing matrix, and the transition moment is related to the
off diagonal one, as shown in Fig. \ref{fg:susy}.
This reaction, along with the companion LFV decay
$\mu^+ \rightarrow e^- \gamma$, will be 
searched for in ``next generation'' experiments now under
construction.\cite{meg,meco}

From neutrino oscillations
we already know that lepton flavor is violated, and this violation will
be enhanced if there is new dynamics at the TeV scale.  This same new 
physics could also generate measurable effects in the magnetic and electric
dipole moments of the muon as well.\cite{LFV,ellis1,ellis2,bdm}

\section{Measurement of the muon anomaly}
The method used in the third
CERN experiment and the BNL experiment are
very similar, save the use of direct muon 
injection\cite{kick} into the storage ring,\cite{mag,inf}
which was developed by the E821 collaboration.  These
experiments are based on the
fact that for  $a_{\mu} > 0$ the spin 
precesses faster than
the momentum vector when a muon travels transversely to a 
magnetic field.  The spin precession frequency $\omega_S$
consists of the Larmor and Thomas spin-precession terms. The
spin frequency $\omega_S$, the momentum   
precession (cyclotron) frequency $\omega_C$,  are given by 
\begin{equation}
 \omega_S = {geB \over 2 m c} + (1-\gamma) {e B \over \gamma mc};\qquad
 \omega_C = {e B \over mc \gamma}.
\label{eq:omegas}
\end{equation}
The difference frequency
\begin{equation}
\omega_a = \omega_S - \omega_C = \left({g-2 \over 2}\right) {eB \over mc}
\label{eq:omeganoE}
\end{equation}
is the frequency with which the spin
precesses relative to the momentum, and is  proportional to
the anomaly, rather than to $g$.
A precision measurement of $a_{\mu}$ requires precision measurements
of the muon spin precession frequency $\omega_a$,  and the magnetic field,
which is expressed as the free-proton precession frequency
$\omega_p$ in the storage ring magnetic field.

The muon frequency can be measured as accurately as the counting
statistics and detector apparatus permit.  
The design goal for the NMR magnetometer and calibration system
was a field accuracy of 0.1 ppm.  The $B$ which enters in 
Eq. \ref{eq:omeganoE} is the average field seen by the ensemble of muons
in the storage ring.  In E821 we reached a precision of 0.17 ppm in the
magnetic field measurement.

An electric quadrupole\cite{quads} is used for vertical focusing, 
taking advantage of the 
``magic''~$\gamma=29.3$ at which an electric field does not contribute to
the spin motion relative to the momentum. With both an electric
and a magnetic field, the spin difference frequency is given by
\begin{equation}
\vec \omega_a = - {e \over mc}
\left[ a_{\mu} \vec B -
\left( a_{\mu}- {1 \over \gamma^2 - 1}\right) \vec \beta \times \vec E
\right],
\label{eq:tbmt}
\end{equation}
which reduces to Eq. \ref{eq:omeganoE} in the absence of an electric field.
For muons with $\gamma = 29.3$ in an electric field alone,
the spin would follow the momentum vector.

\begin{figure}[h!]
  \includegraphics[width=.45\textwidth]{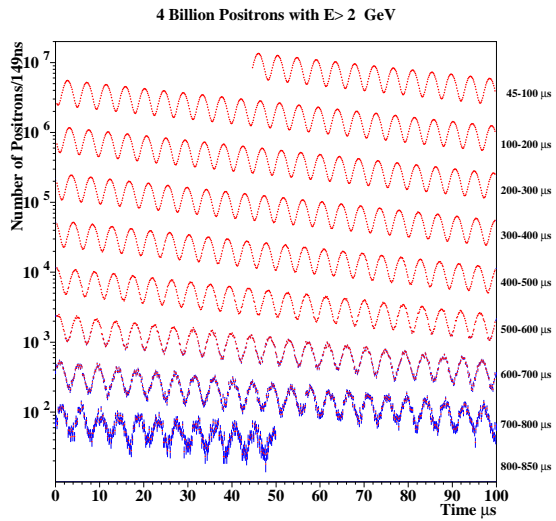}
  \caption{The time spectrum of positrons with energy greater than
2.0 GeV from the year 2000 run.  The endpoint energy is 3.1 GeV.
The time interval for each of the diagonal ``wiggles'' is given
on the right.}
\label{fg:wig00}
\end{figure}

The experimental signal is the $e^{\pm}$ from $\mu^{\pm}$ decay, which 
were detected by lead-scintillating
fiber calorimeters.\cite{det}  The time and energy of each event was
 stored for analysis offline. 
Muon decay is a three-body decay, so the 3.1 GeV muons produce a continuum
of positrons (electrons) from the end-point energy down.  Since the highest
energy  $e^{\pm}$ are correlated with the muon spin, if one counts high-energy 
 $e^{\pm}$ as a function of time, one gets an exponential from muon decay
modulated by the $(g-2)$ precession. The expected form for the positron time
spectrum is $f(t) =  {N_0} e^{- \lambda t } 
[ 1 + {A} \cos ({\omega_a} t + {\phi})] $, however in analyzing the
data it is  necessary
to take a number of small effects into account in order to obtain
a satisfactory $\chi^2$ for the fit.\cite{bennett1,bennett2}
The data from our 2000 running period is shown in Fig. \ref{fg:wig00}

\begin{figure}[h!]
  \includegraphics[width=.45\textwidth]{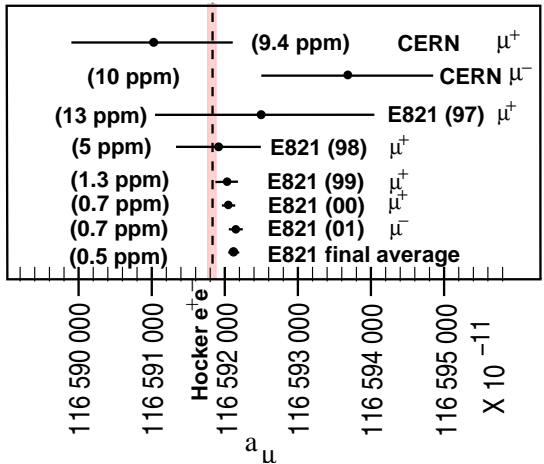}
  \caption{Measurements of $a_{\mu}$. The strong interaction 
contribution is taken from 
references \cite{HMNT} and \cite{HICHEP}. }
\label{fg:amu}
\end{figure}

The experimental results from E821 are shown in Fig. \ref{fg:amu}, with
the average 
\begin{equation}
a_\mu(\rm{E821}) = 11\,659\,208(6) \times 10^{-10}
\qquad (\pm 0.7\ {\rm ppm})
\end{equation}
which determines the ``world average''.  The theory value\cite{HICHEP,HMNT}
\begin{equation}
a_\mu(\rm{SM}) = 11\,659\,182(6) \times 10^{-10}
\qquad (\pm 0.7\ {\rm ppm})
\end{equation} 
is taken from H\"ocker et al.,\cite{HICHEP}, which updates 
their earlier analysis\cite{dehz2} with the KLOE data;\cite{KLOE}
and from 
Hagiwara, et al.,\cite{HMNT} who use a different weighting scheme for
the experimental data when evaluating the dispersion integral 
but do not include the KLOE data.
When this theory value is compared to the standard model value
using either of these two analyses\cite{HICHEP,HMNT}
for the lowest-order hadronic
contribution, one finds
\begin{equation}
\Delta a_{\mu}({\rm E821-SM}) = ( 26 \pm 9.4  )\times 10^{-10}, 
\label{eq:delta}
\end{equation}
or a discrepancy of 2.7 standard deviations.

To show the sensitivity of our measurement of
$a_{\mu}$ to the presence of virtual electroweak gauge bosons, we
subtract off the electroweak contribution of 
$ 15.4 (0.1)(0.2) \times 10^{-10}$
from the standard model value, compare with experiment and obtain
\begin{equation}
\Delta a_{\mu} = (40.7 \pm 8.6)  \times 10^{-10}. \label{eq:deltanw}
\end{equation}
a 4.7 standard deviation discrepancy.  This difference shows conclusively
that E821 was sensitive to physics at the 100 GeV scale.  At present,
it is inconclusive whether 
we see evidence for contributions from physics beyond the standard-model
gauge bosons.

With each data set, the systematic error was reduced, and for the final
data set taken in 2001 the
systematic error on $a_{\mu^-}$ was 0.27 ppm
with a statistical error of 0.66 ppm.
Given the tantalizing discrepancy 
between our result and the latest standard-model
value, and the fact that the hadronic error could be
reduced by about a factor of two over the next few years,\cite{davmar} 
we submitted a
new proposal to Brookhaven to further improve the experimental measurement.
The goal of this new experiment is $\pm 0.2$ ppm total error, with the
goal of controlling the total systematic errors on the magnetic field
and on the muon frequency measurement to 0.1 ppm each.

Our proposal\cite{E969} was given enthusiastic scientific approval in 
September 2004 by the Laboratory, and has been given the new
number, E969. Negotiations are underway between the Laboratory and
the funding agencies to secure funding. 

A letter of intent (LOI) for an even more precise \g2 experiment was also
submitted to J-PARC.\cite{g2jparc}  In that LOI we proposed to reach a
precision below 0.1 ppm.  Since it is not clear how well the hadronic
contribution can be calculated, and whether the new Brookhaven experiment
E969 will go ahead, we will evaluate whether to press forward with this
experiment at a later time.  Our LOI at J-PARC\cite{g2jparc} 
was predicated on
pushing as far as possible at Brookhaven before moving to Japan.

\section{The Muon EDM}

While the MDM has a substantial standard model value, the predicted EDMs
for the leptons are unmeasurably small and lie orders of magnitude below
the present experimental limits given in Table \ref{tb:edm}.
An EDM at a measurable level would signify physics 
beyond the standard model. 
SUSY models, and other dynamics at the TeV scale
do predict EDMs at measurable
levels.\cite{ellis1,ellis2,bdm,fm}

\begin{figure}[h!]
\includegraphics*[width=.45\textwidth]{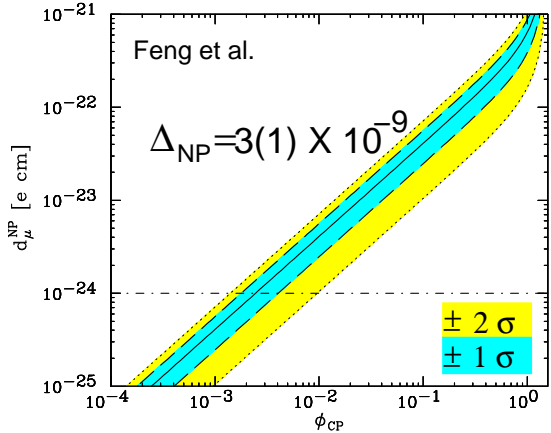}
\caption{The range of  $\phi_{CP} $ available to a dedicated 
muon EDM experiment.\cite{fm}  
The two bands show the one and two standard-deviation
ranges if $a_{\mu}$ differs from the standard model 
value by  $(3\pm 1)\times 10^{-9}$.}
\label{fg:phicp}
\end{figure}

A new experiment to search for a permanent EDM of the muon 
with a design sensitivity of $10^{-24}$ $e$-cm is being
planned for J-PARC.\cite{loi}  This sensitivity lies well within 
values predicted by some SUSY models.\cite{bdm} 
Feng, et al.,\cite{fm} have 
calculated the range of $\phi_{CP} $ available to such an experiment,
assuming a new physics contribution
to $a_{\mu}$ of $3 \times 10^{-9}$,
\begin{equation}
d_{\mu}^{\rm NP} \simeq 3 \times 10^{-22}\left({a_{\mu}^{ \rm NP} \over
3 \times 10^{-9}} \right)
\tan \phi_{CP} \ \ e{\rm -cm}
\end{equation}
where  $\phi_{CP}$ is a {\sl CP} violating phase.  This range is
shown in Fig. \ref{fg:phicp}.

Of course one wishes to measure as many EDMs as possible to understand
the nature of the interaction.  While naively the muon and electron EDMs
scale linearly with mass, in some theories the muon EDM is
greatly enhanced relative to linear scaling relative to the electron EDM
when the heavy neutrinos of the theory are 
non-degenerate.\cite{ellis1,ellis2}

\begin{figure}[h!]
\begin{center}
\includegraphics*[width=.3\textwidth]{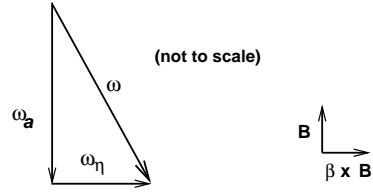}
\end{center}
\caption{A sketch showing the relationship between $\omega_a$ and 
$\omega_{\eta}$.}
\label{fg:omegaeta}
\end{figure}

With an EDM present, the spin precession relative to the momentum is given 
by
\begin{eqnarray}
\vec \omega  &=& 
 -{e \over m} 
\left[ a_{\mu} \vec B -
\left( a_{\mu}- {1 \over \gamma^2 - 1} \right) {{\vec \beta \times \vec E }\over c }
\right]  \nonumber \\
&\qquad \ \  +&
{e \over m}\left[ {\eta \over 2} \left( {\vec E \over c} +
\vec \beta \times \vec B \right) \right] 
\label{eq:omegawedm}
\end{eqnarray}
where
\begin{equation}
d_{\mu} = {\eta\over 2} ({e \hbar \over 2 m c }) \simeq \eta \times 4.7\times 
10^{-14} \ \ e{\rm  - cm}
\end{equation}
and 
$a_{\mu} = ({g-2 \over 2})$.  For reasonable values of 
$\beta$, the motional electric field 
$\vec \beta \times \vec B$ is much larger than electric fields that can be
obtained in the laboratory, and the two vector frequencies are orthogonal
to each other. 

The EDM has two effects on the precession:
the magnitude of the observed frequency is increased, and the
precession plane is tipped relative to the magnetic 
field, as illustrated in Fig. \ref{fg:omegaeta}.
E821 was operated at the magic $\gamma$ so that the focusing 
electric field did not cause a spin precession.  
In E821 the tipping of the 
precession plane is very small, ($\eta/2 a_{\mu} \simeq 9$ mrad)
if one uses the CERN limit\cite{cern3} given in Table \ref{tb:edm}.  
This small tipping
angle makes it very difficult to observe an EDM effect in E821, since
the \g2 precession ($\omega_a$) is such a large effect.

\begin{figure}[h!]
\includegraphics*[height=.45\textwidth]{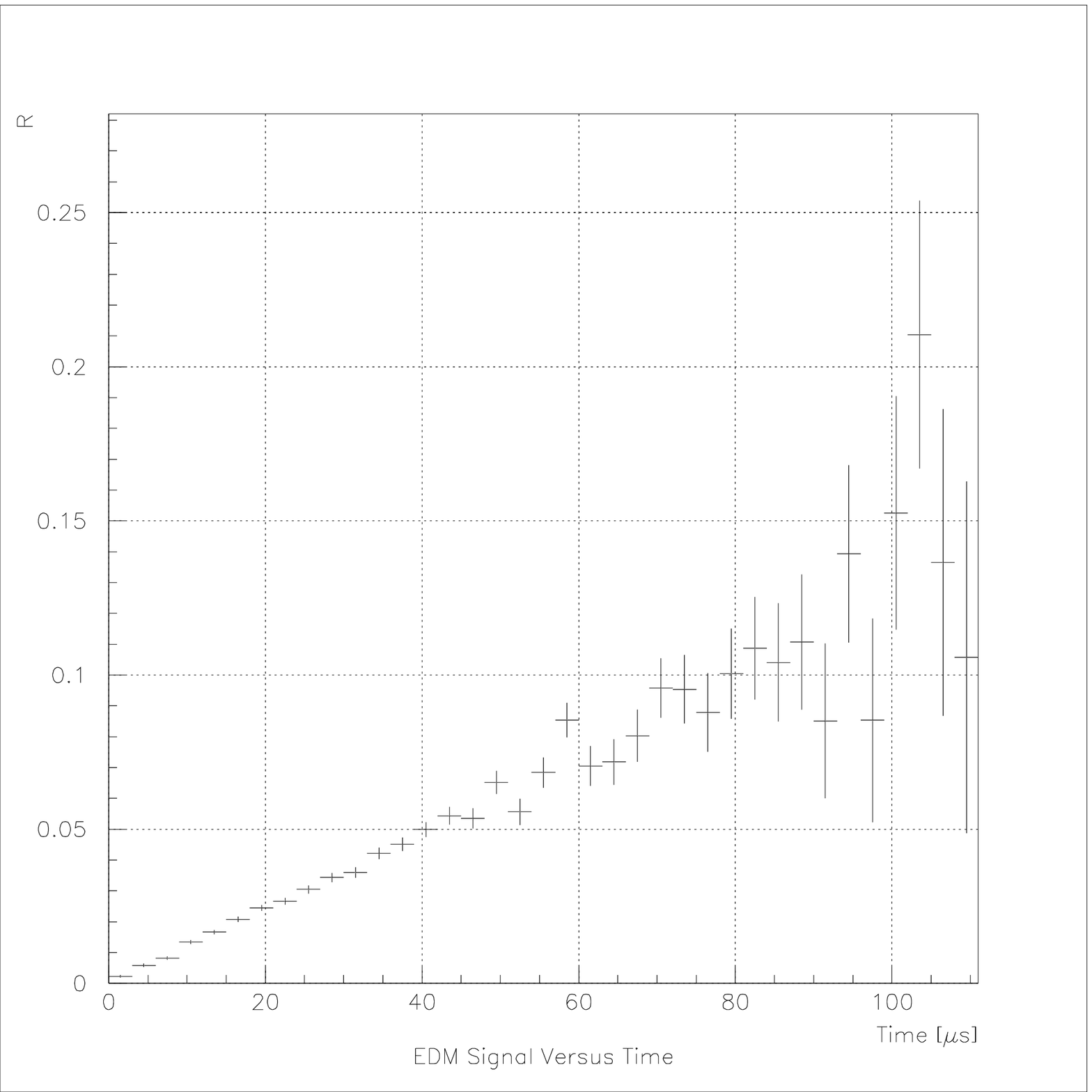}
\caption{Simulation of an EDM signal for \break $\eta = 5 \times 10^{-7}$ or
$d_{\mu} = 2.3 \times 10^{-20}$ }
\label{fg:edmsig}
\end{figure}

We have recently introduced a new idea which optimizes the EDM signal,
and which uses the motional electric field in the rest frame of the muon
interacting with the EDM to cause spin motion.\cite{farley}
The dedicated 
experiment will be operated off of the magic $\gamma$, 
for example at $\sim500$ MeV/c, 
and will use a radial electric field to stop the $(g-2)$ 
precession.\cite{farley}  Then the spin will follow the momentum
as the muons go around the ring, except for any movement arising from
an EDM.  Thus the EDM would cause a steady build-up of the spin out of the
plane with time.  Detectors would be placed above and below the storage 
region, and a time-dependent up-down asymmetry $R$ would be the signal of
an EDM, 
\begin{equation}
R = { N_{\rm up} - N_{\rm down} \over N_{\rm up} + N_{\rm down}}.
\end{equation}
A simulation for $d_{\mu}= 2\times 10^{-20}\ e$ cm is given in 
Fig. \ref{fg:edmsig}.

The figure of merit for statistics in the EDM experiment is the number
of muons times the polarization.
In order to reach $10^{-24}\ e$ cm, 
the muon EDM experiment would need $NP^2 \simeq 5 \times 10^{16}$,
a flux only available at a future facility.  While progress can
still be made at Brookhaven on $a_{\mu}$, a dedicated muon EDM experiment
must be done elsewhere.

\section{Summary and Conclusions}

Measurements of the muon and electron anomalies played an important
role in our understanding of sub-atomic physics in the 20th century.
The electron anomaly was tied closely to the development of QED.
The subsequent measurement of the muon anomaly showed that the muon
was indeed a ``heavy electron'' which obeyed
QED.\cite{cern3}  With the sub-ppm accuracy now available for the 
muon anomaly,\cite{brown2,bennett1,bennett2} 
there may be indications that new physics is beginning to
appear in loop processes.\cite{mar}
 
The non-observation of an electron EDM 
is becoming an issue for supersymmetry, just as the non-observation
of a neutron EDM implies such a mysteriously (some would say
un-naturally) small 
$\theta$-parameter for QCD. The search for EDMs will continue, and
if one is observed, the motivation for further searches in other systems
will be even stronger.  The muon presents a unique opportunity
to observe an EDM in a second-generation particle, where the {\sl CP}
phase might be different from the first generation, or the scaling
with mass might be quadratic rather than linear.

In closing I consider two scenarios: (i) the LHC finds SUSY; and 
(ii) the LHC finds the standard model Higgs at a reasonable mass
and nothing else.

If SUSY turns out to be {\it the} extension to the standard model, then
there will be SUSY enhancements to $a_{\mu}$ to the muon EDM and also
to the amplitudes for lepton flavor violating muon decays.
Once the SUSY mass spectrum is measured, $a_{\mu}$ will provide a
very clean measurement of $\tan \beta$.\cite{davmar}

If SUSY or other new dynamics at the TeV scale are not found at LHC,
then precision experiments, which are sensitive through virtual loops
to much higher mass scales than direct searches for new particles,
become even more important.  Experiments 
such as EDM searches, \g2 and searches for 
lepton flavor violation, all carried out at high intensity
facilities, may provide the only way to probe these higher energy scales.

Opportunities at future high intensity facilities are actively being
pursued, and both the theoretical and experimental situations are evolving.
It is clear that the study of lepton moments and lepton flavor
violation, along with  neutron
EDM searches, will continue to be 
a topic of great importance in the first part of the 21st century.

{\em Acknowledgments: 
I wish to thank my colleagues on the muon \g2 experiment,
 as well as  
M. Davier, J. Ellis,  E. de Rafael,  W. Marciano and T. Teubner  for
helpful discussions. Special thanks to Y. Semertzidis for 
critically reading this manuscript.
}




\begin{thebibliography}{9}
\bibitem{gab} G. Gabrielse and collaborators have constructed a new trap, and
expect to improve on the electron anomaly by a factor of 10 to 100 over
the measurements of \cite{eg2}
\end{thebibliography}


\end{document}